\documentclass[10pt,graphicx,subfigure,axodraw,cleverref]{paper}
\usepackage{bm}
\usepackage{multirow}
\usepackage{amsfonts}
\usepackage{amssymb}
\usepackage{epsfig}
\usepackage{float}
\usepackage{ulem}
\restylefloat{table}
\usepackage[usenames,dvipsnames]{color}
\newcommand{\beqn}{\begin{eqnarray}}
\newcommand{\eeqn}{\end{eqnarray}}
\newcommand{\be}{\begin{equation}}
\newcommand{\ee}{\end{equation}}
\newcommand{\non}{\nonumber \\}

\newcommand{\cl}{{\cal L}}

\def\st{Stueckelberg~}
\def\s1{$s_{\alpha}$}
\def\s2{$s_{\gamma}$}
\def\s3{$s_{\delta}$}
\def\c1{$c_{\alpha}$}
\def\c2{$c_{\gamma}$}
\def\c3{$c_{\delta}$}

\def\45{\overline{45}}
\def\5{\overline{5}}
\def\70{\overline{70}}
\def\50{\overline{50}}

\def\non{\nonumber\\}

\def\c{\acute{c}}

\def\45{\overline{45}}
\def\5{\overline{5}}
\def\70{\overline{70}}
\def\50{\overline{50}}

\usepackage{amsmath,amsfonts,amssymb}
\usepackage{hyperref}
\usepackage{cleveref}
\crefname{equation}{Eq.}{Eqs.}
\crefname{figure}{Fig.}{Figs.}
\crefname{table}{Table}{Tables}
\crefname{section}{Section}{Sections}

\usepackage{epsfig}
\usepackage{psfrag}
\usepackage{dsfont}

\def\GeV{\rm GeV~}

\def\cl{{\cal L}}

\newcommand{\beq}{\begin{equation}}
\newcommand{\eeq}{\end{equation}}

\def\sugra{supergravity ~grand~ unification~}
\def\gut{grand~unification~}

\def\s{\scriptsize}

\begin{document}

\baselineskip 18pt
\begin{titlepage}

\begin{flushright}
\end{flushright}

\begin{center}
%\fbox{\bf{\today}}
{\bf {\ {\textsf{\Large{
% {\boldmath}
    }}}}}

\vskip 0.5 true cm \vspace{0cm}
\renewcommand{\thefootnote}
{\fnsymbol{footnote}}

%\fbox{\bf{\today}}\\~\\

{\bf \Large   High Energy Physics and Cosmology  at the Unification Frontier:
Opportunities and Challenges in the coming years}\\~\\
Pran Nath\footnote{Email: p.nath@notheastern.edu} %\skip 0.5 true cm}
\end{center}
\begin{center}
{\noindent
\textit{Department of Physics, Northeastern University,
Boston, MA 02115-5000, USA}   }\\

\end{center}

\vskip 1.0 true cm \centerline{\bf Abstract}

We give here an overview of recent developments in high energy physics and cosmology and their interconnections 
 that relate to  unification, and  discuss  prospects for the future.
 Thus there  are currently three empirical data that point to supersymmetry as an underlying symmetry of particle physics: the unification of gauge couplings within
  supersymmetry, the fact that nature respects the supersymmetry prediction that the Higgs boson mass lie 
  below 130 GeV, and vacuum stability up to the Planck scale with a Higgs boson mass at $\sim 125$ GeV 
  while the standard model does not do that. Coupled with the fact that 
 supersymmetry solves the big hierarchy problem related to the quadratic divergence to the Higgs boson mass square
 along with the fact that  there is no alternative paradigm that
  allows us to extrapolate physics from the electroweak scale to the grand unification scale consistent with
  experiment, supersymmetry remains a compelling framework for new physics beyond the standard model.
 The large loop correction to the Higgs boson mass in supersymmetry to lift the tree mass to the experimentally
 observable value, indicates a larger value of the scale of weak scale supersymmetry, making the observation of
 sparticles more challenging but still within reach at the LHC for the lightest ones. Recent analyses show that
 a high energy LHC (HE-LHC) operating at 27 TeV running  at its optimal luminosity of $2.5 \times 10^{35}$ cm$^{-2}$s$^{-1}$
 can reduce the discovery period by several years relative to HL-LHC and significantly extend the reach in parameter space
 of models. In the coming years several experiments related to neutrino physics, searches for supersymmetry, on dark matter
 and  
dark energy will have direct impact on the unification frontier.  Thus the discovery of sparticles will establish supersymmetry 
as a fundamental symmetry of nature and also lend direct support for strings.  Further, discovery of sparticles associated
with missing energy will constitute discovery of dark matter with LSP being the dark matter.
On the cosmology front 
 more accurate measurement of the equation of state, i..e, $p= w \rho$, will shed light on the 
 nature of dark energy. Specifically, $w> -1$  will likely indicate the existence of a dynamical field, possibly quintessence, 
 responsible for dark energy and $w<-1$ would indicate an entirely new sector of physics.
   Further, more precise measurements of the ratio $r$ of tensor to scalar
 power spectrum, of the scalar and tensor spectral indices $n_s$ and $n_t$ and of non-Gaussianity will hopefully
 allow us to realize a standard model of inflation. These results  will be a guide to further model building 
 that incorporates unification of particle physics and cosmology.


\medskip
\noindent

\end{titlepage}
\date{today}

{\bf 
\tableofcontents
}

\section{Introduction\label{sec1}}	
The  standard model of particle physics~\cite{sm}
describing the electro-weak and the strong 
interactions is a  remarkably successful model up to the TeV scale. With the discovery of the Higgs boson 
the spectrum of the standard model is now complete ~\cite{Englert:1964et,Chatrchyan:2012xdj}.
However, because of the hierarchy
problem ~\cite{Gildener:1976ih}, 
one expects new physics beyond that scale. Supersymmetry provides a simple mechanism that
cancels the large loop corrections to the Higgs boson that arise from fermions  that couple to  the Higgs 
boson and is currently the leading candidate for new physics beyond the standard model. 
 Initially supersymmetry appeared in two dimensions  and later promoted to four dimensions
relevant for phenomenology~\cite{early-susy}.
However, if supersymmetry is to be a fundamental supersymmetry it must be a local symmetry~\cite{sugra1,sugra2}  
    (For a more extensive discussion
    see ~\cite{Nath:2016qzm}). 
  This means grand unified theories~\cite{gut-models} which unify the electro-weak and the strong interactions cannot just be 
  globally supersymmetric extensions of the grand unification~\cite{Dimopoulos:1981zb}  
  but must be locally supersymmetric grand unified models\cite{msugra}. 
   Supergravity \gut is not a quantum theory of gravity and  for that we must look to strings.
   While much
 progress has occurred in strings it is still  work in progress and  the possible number 
of  vacuum states of the theory is enormous.  So most of the discussion in this paper will focus on supersymmetry and supergravity 
based models. 
A unified model of particle physics has implications along many frontiers of fundamental physics:  
the energy frontier, the intensity frontier  and the cosmic frontier.
 There is, of course,  a significant overlap among the three frontiers and particle physics plays a role
 in all. 
 
The outline of the rest of the paper is as follows: In section (\ref{sec2}) we discuss the energy frontier, in section (\ref{sec3})
the intensity frontier, and in section (\ref{sec4})  the cosmic frontier.  Unification frontier is discussed in section (\ref{sec5}).
Challenges for high energy physics and cosmology are discussed in section (\ref{sec6}) and future high energy colliders in section (\ref{sec7}). 
Conclusions are given in section (\ref{sec8}).  Notation and some basic formulas are given in section (\ref{sec9}).

\section{The Energy Frontier \label{sec2} }
Supersymmetry has not been observed so far.
The explanation of this is easily found in the measured value of the Higgs boson mass at $\sim 125$ GeV.
Thus at the tree level the Higgs boson  mass is less than $M_Z$ which tells us that the loop correction 
must be large to bring the mass up to $\sim 125$ GeV observed by experiment. 
Remarkably  in SUGRA models the Higgs boson mass is predicted so that 
$M_H \leq 130~{\rm GeV}$~\cite{higgs-sugra-prediction}.
 In fact, the measurement of the Higgs  boson at $125$ GeV  gives further support for supersymmetry~\cite{Nath:2017lma}.
  This is in part because of vacuum stability.
 Thus for  large field configurations where $h>> v$ the Higgs potential is governed by the quartic term
$ V_h\sim \lambda_{eff} h^4$.  For vacuum stability we need $\lambda_{eff}>0$ up to the Planck scale. 
It turns out that  vacuum stability depends critically on the top mass and a 
 larger top mass makes the vacuum more
unstable.   
An advanced precision analysis  including two-loop matching, three-loop renormalization group evolution, and pure QCD corrections through four loops  
   gives an upper bound on the top pole mass  of $ m_t^{\rm cri} = (171.54 \pm0.30^{+0.26}_{-0.41}) {\rm ~GeV}$  for SM stability up to the Planck  scale. On the other hand the experimental value of the top mass is $ m^{\rm exp}_t= (173.21\pm 0.51 \pm0.71) {\rm ~GeV}$
   which  makes the vacuum stable only up to about  $10^{10}-10^{11}$ GeV. In models based on supersymmetry with a Higgs mass of $125$ GeV, the  vacuum can be stable up to the Planck scale~\cite{vac-stability}.   

 One might ask what  light the Higgs boson mass might shed on the nature of soft parameters.  In supergravity models with universal 
 boundary conditions the soft parameters are described by $m_0, m_{1/2}, A_0, \tan\beta, {\rm sign}(\mu)$, where 
 $m_0$ is the universal scalar mass, $m_{1/2}$ is the universal gaugino mass, $A_0$ is the universal trilinear coupling, 
 $\tan\beta$ is the ratio $<H_2>/<H_1>$ where $H_2$ gives mass to the up quarks, and $H_1$ gives mass to the down quarks
 and leptons. We note in passing  that supergravity models allow for non-universalities:  in 
 the scalar masses, in the gaugino mass masses and in the trilinear couplings. For simplicity we consider  the universal case.
 Here an analysis within  supergravity grand unification \cite{Arbey:2012dq}
 shows that to get the Higgs boson mass consistent with experiment one needs a large $A_0$ which is easy to produce
 in supergravity grand unified models. 
 We note that 
 there are two likely reasons why SUSY has not been observed so far at the LHC.
 The first one is is that as mentioned  the Higgs boson mass measured at 125 GeV points  to the weak SUSY 
 scale to be large lying in the several TeV range which makes the sparticles less accessible. 
  Second,  typically in large weak SUSY scale models the neutralino is mostly a Bino and one needs  
 coannihilation to satisfy the relic density constraint as given by the PLANCK experiment~\cite{Ade:2015xua}.
 Coannihilation, of course,  requires a compressed sparticle spectrum which makes the observation of a supersymmetric signal
 difficult at the LHC.  We note that the above discussion relates to analyses based on supergravity grand unification and the 
 analyses for gauge mediation~\cite{gauge-mediation} and anomaly mediation~\cite{anomaly-mediation} are  different.

The large scale of weak scale supersymmetry lying in the several TeV region raises the issue of naturalness
since the electroweak scale is related to $\mu$ and  the soft parameters through radiative breaking of the electroweak symmetry. 
For perspective we note that there are various naturalness problems in particle physics. The most prominent one is that of the cosmological constant.
Data from supernovae cosmology project indicates that $\Lambda$ that enters in the Einstein equations has the value
$\Lambda\sim 4.33 \times 10^{-66}$ eV$^{2}$. However, theoretically loop corrections to the $\Lambda$  could be as large as $M_{Pl}^2$
which makes the theoretical prediction a factor of 120 orders of magnitude larger than the experimentally observed value. 
Another, example of a larger naturalness problem is the loop correction to the Higgs boson mass square in the standard model. 
 Here the loop correction is larger by a factor of $10^{28}$. In this case supersymmetry provides a solution.
 There are a variety of other smaller hierarchies where factors of up to few (i.e., 2-3) orders of magnitude are
 involved. Such small hierarchies could often be due to a partial knowledge of the boundary conditions of the problem. 
  One such case is the little hierarchy problem related to the electroweak symmetry breaking. Here the electroweak symmetry breaking 
  relation involves soft parameters such as $m_0, m_{1/2}, A_0, \tan\beta$, the $\mu$ parameters and the mass $M_Z$ of the $Z$ boson.
  This has led to various criteria regarding how large the soft parameters can be relative to the electroweak scale. Some of the early
  criteria were based on examining only one branch of radiative breaking which we will call the Euclidean branch (EB). Further, in the 
  pre-LHC era, one imagined that the Higgs boson mass was likely to be only modestly above the 
tree level mass, which would imply a low SUSY scale and 
  led to  the expectation of an early discovery of supersymmetry. However, as mentioned already  
  the discovery of the Higgs boson at 125 GeV, points up the
  existence of a large loop correction to the tree value of the Higgs and thus a correspondingly much higher value of the SUSY scale.\\

Regarding the  large scale of weak scale supersymmetry,  it was shown quite a while ago that such scales  can be natural. 
Thus in the analysis of electroweak symmetry breaking (for a review see
\cite{Ibanez:2007pf}) it was shown 
that TeV size scalars can arise quite naturally on the 
hyperbolic branch ~\cite{hb}.
To see this more clearly let us look at Eq. (\ref{rewsb1}) which involves the soft parameters 
 $m_0, m_{1/2}, A_0, \tan\beta$, and the Higgs mixing parameter $\mu$, 
where  the relation holds at scale $Q$ to one loop order and where  $\Delta^2_{\text loop}$ is the loop correction.
There is an underlying geometry to the radiative breaking equation Eq. (\ref{rewsb1}).  This underlying geometry can be more
easily seen by choosing  a renormalization group scale $Q$ where the loop correction to the tree equation is minimized.
 One then  finds  the following phenomenon: the sum of the terms which involve $A_0$ and $m_{1/2}$ sum up 
  to be positive while $C_1$ has an ambiguous sign. In certain part of the parameter space, $C_1$ is positive.
  In this case all the terms on the right hand side of Eq. (\ref{rewsb1})  
   are positive, and thus if we were to fix $\mu$ there
  would be an upper bound to $m_0$, $A_0$ and $m_{1/2}$. This is the elliptical branch of the radiative breaking of the electroweak
  symmetry. This was the only branch known~\cite{Barbieri:1987fn}  
   till the analysis of~\cite{hb}  
 and led to 
  various criteria for fine tuning and implication that naturalness implies low values for the soft parameters. 
  
  However, in the analysis of~\cite{hb}  
    a new branch of radiative breaking was discovered where one finds that one, two or all three soft parameters could be 
    large while $\mu$ remains small.
   This is the hyperbolic branch of radiative breaking of the electroweak symmetry. The hyperbolic branch contains 
   focal points, focal curves and focal surfaces as discussed in section \ref{sec9.1}. 
      For the case when two soft parameters such as $m_0$ and $A_0$ get large while $\mu$ is small and fixed, one has a focal curve.
    When all soft parameters get large while $\mu$ remains small and fixed, one has a focal surface. For the case when
    the front factor in the term with $m_0^2$ vanishes, one has a focal point as explained in section \ref{sec9.1}.   
     Thus for all cases, i.e.,  the focal curve, the  focal surface,  and the focal point 
 one may  have a relatively small $\mu$ while the weak scale lies in the several TeV region. 
More recent analyses show that squark masses can lie in the 50-100 TeV region
consistent with radiative breaking of the electroweak symmetry, the Higgs boson mass constraint and with  gauge coupling unification, 
which still
allows for some light sparticles to be accessible at LHC energies ~\cite{Aboubrahim:2017wjl}.  

We note in passing that a high value of the scale of weak scale supersymmetry has some important benefits.
Thus a high value of the weak supersymmetry scale resolves the problem of gravitino decay, in that in supergravity models 
 a decay of the gravitino with mass  in the 10 TeV and above region 
would not upset the BBN and  would only minimally contribute to the relic density for up to reheat temperatures of $10^{12}$ GeV in the 
post inflationary period.  A high SUSY scale also helps suppress contributions from the SUSY CP phases to the  edm of the quarks and the 
leptons (see, e.g.,~\cite{Ibrahim:2007fb}). Further,  it helps suppress proton decay from the lepton and baryon number violating dimension five 
operators~\cite{Nath:2006ut,Liu:2013ula}.

       In most models with high scale of weak scale supersymmetry, the neutralino turns out to be mostly a bino. 
       However, as  discussed earlier 
         the annihilation 
       cross section of bino's in the early universe is rather small,  and one needs an additional boost from co-annihilation to be consistent
       with the relic density constraint on neutralino dark matter (see, e.g., ~\cite{coann}).       
      Further, as noted earlier coannihilation, of course,  leads to a compressed spectrum  and a compressed spectrum 
can appear in models with universal soft breaking as well as in models with 
 non-universal soft breaking at the grand unification scale (for recent works see~\cite{compressed}).  
 Such non-universalities can be in the gaugino sector,
 in the Higgs sector, and in the sfermion sector.
 The existence of these non-universalites leads to a large landscape in the
 sparticle mass spectrum~\cite{sparticle-landscape}.
Even though the scalar masses are large one could still observe supersymmetric signals if the gaugino masses are relatively small. 
In this case the lightest sparticles would be the neutralino, the chargino, the gluino and possibly a light stau and a light stop. 
For experimental searches for supersymmetry with a compressed spectrum and associated works 
see~\cite{compressed-search,Nath:2010zj}.

 In addition to the search for sparticles, signatures of supersymmetry might arise from the observation of additional Higgs bosons.
Thus in supersymmetry and supergravity based models, one needs at least on pair of Higgs doublets, which after electroweak symmetry
breaking will lead to additional Higgs bosons  beyond the lightest Higgs boson $h^0$ which include the heavier CP even Higgs $H^0$,
a pseudo-scalar Higgs boson $A^0$ and a charged Higgs boson $H^{\pm}$. It is known that the presence of CP phases for the soft 
parameters can induce a mixing of the CP even and the CP odd Higgs bosons, i.e., mixing among $h^0, H^0, A^0$ which can produce
observable effects at accelerators~\cite{cp-mixing}.
Such effects may be visible  if the heavier neutral Higgs bosons become accessible at the
energy and luminosity of the HL-LHC.   
Further, in addition to   the extra Higgs states that appear in supersymmetric models, there are a variety of competing models such 
as the  two Higgs doublet model, twin Higgs and many others (for some recent works on alternative Higgs
models see, e.g.,   \cite{Gunion:1989we,Chacko:2005pe,Csaki:2017cep}, 
 and for recent Higgs related works see \cite{higgs-related}).

\section{Intensity Frontier \label{sec3}} Precision measurements in 
 high intensity experiments can point to new physics beyond the standard model
 (for a comprehensive review see ~\cite{Hewett:2012ns}).
 These new phenomena may be manifest via rare processes and via loop  corrections that go beyond those predicted by the standard model.
 Thus, for example, the 
 existence of new sources of CP violation would be indicated by the detection of edms of quarks and leptons, a topic which will be discussed
 in section \ref{sec3.2}.
A deviation in $g_\mu-2$ from the standard model prediction if confirmed would also indicate the existence of new physics.
Regarding  the Brookhaven (BNL) experiment E821, carried out in the period 1999-2001, 
  a $\sim 3.5\,\sigma$ deviation from the standard model prediction was seen so that 
$\delta a_\mu = (287 \pm 80.)\times10^{-11}$ ~\cite{g-2-exp}.
It is known that supersymmetry can produce a significant 
contribution if the smuon, muon-sneutrino and the weak gauginos have masses in the few 
hundred GeV region~\cite{Kosower:1983yw,Yuan:1984ww,g-2-tanbeta}.
On the other hand one expects the sfermions to be heavy in view of the Higgs boson mass.
These results can be reconciled if we assume  $m_3>>m_1, m_2, m_0$. In this case the high gluino mass 
drives the squark masses to be heavy but leaves the slepton masses and weak gaugino masses  to be low 
and it is possible to have consistency with the Brookhaven result~\cite{Akula:2013ioa}.
There are a variety of other models which can also achieve consistency with the data (see, e.g., ~\cite{g-2-vector}).
A new measurement of the muon anomalous moment experiment E989  is underway by the  Fermilab Muon $g-2$ Collaboration.
The Fermilab experiment will accumulate about 21 times more data than the Brookhaven experiment!\cite{Gohn:2017dsp}.
Additionally uncertainties in the theoretical analyses are also expected to be reduced due to hadronic contributions using
lattice gauge analyses and using more accurate determination of the lepton scattering cross-sections~\cite{Blum:2016lnc}.
Runs on E989 have already started and will continue through 2020. It is expected that the new data will reduce the 
uncertainties by a factor of 4. Further, if the BNL measurement of the central value holds, the improved measurement
could lead to a $7\sigma$ discrepancy and provide a clear signal of new physics~\cite{Gohn:2017dsp}.

  \subsection{Neutrino physics} 
     Neutrinos  are the least massive of the elementary particles and are likely connected to a high scale, such as the 
grand unification scale through the see-saw mechanism~\cite{Mohapatra:2005wg}.
      They have other remarkable properties 
    such as they exhibit oscillations where  one flavor state converts into another. They also exhibit the 
     phenomenon that the matrix $U_\nu$ that takes us from the flavor states  $(\nu_e,\nu_\mu,\nu_\tau)$ to the mass
     diagonal states $(\nu_1,\nu_2,\nu_3)$, i.e.,
 $(\nu_e,\nu_\mu,\nu_\tau)^T = U_\nu(\nu_1,\nu_2,\nu_3)^T,$
  has mixing angles which are very different from those in the quark sector.  Thus $U_\nu$ is parameterized by three 
  mixing angles $\theta_{12}$, $\theta_{13}$ and $\theta_{23}$ and  the  CP phase  $\delta_{CP}$.
In addition one has two Majorana phases $\delta_{21}, \delta_{31}$ 
   which do not enter in the oscillation analysis  (see, e.g., ~\cite{Ibrahim:2007fb} for definition of the mixing angles and phases).
  Let $m_1, m_2,m_3$ be three mass eigenvalues. The neutrino oscillations depend only 
on  the  mass squared differences  $\Delta m^2_{21} \equiv m^2_2 - m^2_1$ and 
$\Delta m^2_{32} \equiv m^2_3 - m^2_2$. Experimentally  solar and neutrino experiments
determine the quantities  $\Delta m^2_{21}$ and $\theta_{12}$ while 
atmospheric neutrino 
and accelerator neutrino experiments
determine $|\Delta m^2_{32}|$ and \(\theta_{23}\).
Currently one has~\cite{Gonzalez-Garcia:2014bfa}   $\sin^2\theta_{12} = 0.307^{+0.013}_{-0.012}$, $\sin^2\theta_{23}= 0.538^{+0.033}_{-0.069}$, $\sin^2\theta_{13}= 0.02206\pm
{0.00075}$ and $\delta_{CP}= (234^{0.43}_{-31})^0$.  
   Regarding the mass-squared differences one has  two possibilities~\cite{Gonzalez-Garcia:2014bfa}   
    (1)  $\Delta m^2_{21}= + 7.6 \times 10^{-5}$ eV$^2$, ~~ $\Delta m^2_{32}= + 2. 4 \times 10^{-3}$ eV$^2$  and   
   (2) $\Delta m^2_{21}= + 7.6 \times 10^{-5}$ eV$^2$, ~~ $\Delta m^2_{32}= - 2. 4 \times 10^{-3}$ eV$^2$. 
   Case (1) is the so called normal hierarchy while case (2) is  the inverted hierarchy. Establishing the correct mass hierarchy
   along with  determining the CP phases will be of significance in constraining grand unified models. 
   Hyper-K experiment using  J-PARC muon neutrino beam~\cite{Abe:2014oxa}   
    along with experiments such as   
   The Long Baseline Neutrino Experiment~\cite{Goon:2012if} in the US will lead the effort in this regard.

   Neutrino oscillations provide us information only on the neutrino mass difference $m_j^2-m_l^2$, and the determination of
    the absolute masses of the neutrinos is a challenge. The neutrino less double beta decay is one available means by which 
    we can fix the absolute neutrino mass. Neutrino mass limits on the sum of the neutrino masses can be gotten 
    by using large-scale power spectrum data  of galaxy surveys~\cite{Cuesta:2015iho}    
      which gives $\sum m_{\nu_i} <0.13$ eV.  It is expected that data from  CMB-S4\cite{Abazajian:2016yjj}      
      and data from the Large Synoptic Survey Telescope (LSST)~\cite{Abell:2009aa}      along with other astrophysical data 
      will allow one to put an upper limit on the sum of $0.03$ eV~\cite{Mishra-Sharma:2018ykh}.         
        The study of  Cosmic Infrared Background provides us  with an alternative way to fix the absolute value of the neutrino mass
    from the radiative decay of a neutrino to one with a lower mass. 
    Thus  a neutrino can decay radiatively to neutrinos with lower masses and  for the 
    neutrino mass eigenstates $\nu_1$,  $\nu_2$,  $\nu_3$, with $m_{\nu_{3}}> m_{\nu_2}  > m_{\nu_1}$ one can have radiative decays so that $\nu_3 \to \nu_1 \gamma, \nu_2\gamma$.
    The reason the radiative decays provide us with a way  to measure the absolute mass of the neutrino is the following:
    Suppose we consider the
decay $\nu_j\to \nu_l\gamma$. In the rest frame of the decay of $\nu_j$ the photon energy is given by
 $E_{\gamma} = (m_j^2- m_l^2)/ (2 m_j)$. Since neutrino oscillations provide us with the neutrino mass difference $m_j^2-m_l^2$, a measurement of the photon energy allows a determination of  $m_j$. 
 Recent analyses from the study of CMB spectral distortions give a lifetime for the decay  for $\tau_{12} \geq 4\times 10^{21}$s
 for the smaller mass splitting and 
$\tau_{13}\sim \tau_{23} \geq 10^{19}$s for the larger mass splitting for the normal hierarchy 
 and similar limits within a factor of few for the inverted hierarchy~\cite{Aalberts:2018obr}.
  The radiative decay  lifetime of the neutrino in the standard model is very large (for a review see ~\cite{Studenikin:2018vnp}). 
However, much lower lifetimes for the 
neutrino decays  can be achieved when one goes beyond the Standard Model ~\cite{large-nu-moments,Aboubrahim:2013gfa}.

\subsection{CP  violation in supersymmetry, strings and branes\label{sec3.2}}
Aside from direct searches for new physics with colliders it is possible to explore new physics up to the 
PeV scale using  the electric dipole moments of elementary particles specifically of the leptons. 
This is so because in the standard model, for example, the electric dipole moment (edm) of  the electron is estimated to be 
$|d_e|\simeq  10^{-38}$ $e$cm
and for the neutron the value lies in the range ($10^{-31}-10^{-33}$) $e$cm 
 (assuming that the QCD phase
$\theta$ is eliminated using the Peccei-Quinn symmetry for the case of the neutron). These limits are 
far beyond the  sensitivity of experiments to measure in the foreseeable feature. However, models 
of new physics beyond the standard model can produce edms which are much larger and of sizes
measurable by the current and improved future experiments. Thus a measurement of the edms of the
fundamental particles will provide a window to new physics beyond the standard model (for a review see~\cite{Ibrahim:2007fb}).
 In supersymmetric theories, there are new sources of CP violation beyond those in the standard model 
 and these phases can be large. In fact such phases typically produce edms for the leptons and for 
 quarks which are significantly in excess of the current experiment if one uses a  sparticle spectrum in the
 sub TeV region. Various procedures have been advocated in the past to overcome this problem including 
 mass suppression by use of heavier masses~\cite{Nath:1991dn} and the cancellation mechanism where contributions from 
 different sources cancel~\cite{cancellation}.
  In the post Higgs boson discovery era one finds that the  sparticle spectrum, specifically
 the scalar masses could be large which then provides an automatic window for the exploration of new physics up to the PeV 
 region using the edms (For a discussion of PeV scale in the context of supersymmetry in previous works see, e.g., \cite{Wells:2004di}).
  Thus suppose the supersymmetric  CP phases are large even maximal, then the experimental
 limit on the edms can be used to constrain the sparticle masses.  This idea has been pursued by a number of authors
~\cite{PeV-scale}

 To make the discussion more concrete let us consider the edm of the electron which is the most stringently 
 constrained one among the edms of the elementary particles. Thus the measurement of the electron edm by 
 the ACME Collaboration~\cite{Baron:2013eja} using the polar molecule thorium monoxide (ThO) gives 
%\beqn
$d_e =(-2.1 \pm 3.7_{\rm stat} \pm 2.5_{\rm syst}) \times  10^{-29} e{\rm cm}$, which  corresponds to an upper limit of 
$ |d_e| < 8.7\times  10^{-29} ~e{\rm cm}\,,$  at 90\% CL. 
The corresponding upper limits on the edm of the muon and  on the tau lepton are not that stringent, i.e., one has 
~\cite{pdg}
$ |d_\mu| < 1.9 \times 10^{-19} ~e{\rm cm} $ for the muon edm and 
 $ |d_\tau| <  10^{-17} ~e{\rm cm}\ ,$ for the tau edm.
   In the future  the sensitivity    
    limit on all of them are expected to increase. Thus for the electron edm   $d_e$  the future 
projections are $|d_e| \lesssim 1 \times 10^{-29} e{\rm cm}$ using Fr \cite{Sakemi:2011zz}, 
$ |d_e| \lesssim 1 \times 10^{-30} e{\rm cm}$ using  {\rm YbF ~molecule} \cite{Kara:2012ay}, and 
$ |d_e| \lesssim 1 \times 10^{-30} e{\rm cm}$ using  WN ~ion \cite{Kawall:2011zz}. 
Similar considerations apply to the neutron edm. The current experimental limit on the edm of the neutron is~\cite{Baker:2006ts}
$|d_n| < 2.9 \times 10^{-26}   e{\rm cm} ~~~(90\% ~{\rm CL})$ and  higher sensitivity is expected from experiments in the future~\cite{Ito:2007xd}.
Such analyses can also be utilized  for the discovery of vector like generations through their mixings with the first three generations. 
Vectorlike generations arise in  a variety of grand unified models,  and in string and in D brane models and have 
 been considered by several authors and their discovery
  would constitute new physics (see, e.g.,\cite{vectorlike}).  Aside from the above, the SUSY CP phases  enter in a variety of low energy 
  phenomena such as  event rates in the direct detection of dark matter \cite{Falk:1998pu,Chattopadhyay:1998wb}, on  flavor changing photonic decay of the b quark~\cite{Gomez:2006uv}  and on many others.

\subsection{Hidden sectors in supergravity, in strings and in branes}
 Hidden sectors arise in supergravity models and in models based on string theory. 
 Thus in supergravity models supersymmetry is broken  in the hidden sector and communicated to the 
 visible sector by gravity mediation.
By a hidden sector one means that the 
matter and gauge fields in the hidden sectors are singlets of the standard model gauge group so there is 
no direct communication between them. However, as mentioned there can be communication between 
them by gravity mediation which means by Planck suppressed operators. The hidden sectors can also be
made visible by $U(1)$ probes where the $U(1)_Y$ of the visible sector
mixes with the $U(1)$ of the hidden sector by  kinetic mixing~\cite{holdom}
and by 
 Stueckelberg mass mixing~ \cite{stueckelberg}.  Extra $U(1)$'s factors arise  in grand unified theories, in strings and in D-branes.
 Thus D brane constructions  typically start with a stack of n branes which has a $U(n)$ gauge symmetry. 
 Since $U(n)\supset SU(n)\times U(1)$,
one has $U(1)$ factors appearing. For example, to construct the standard model gauge
group in D-brane models 
one starts with the gauge groups $U(3)\times U(2)\times U(1)^2$ which results in
$SU(3)\times SU(2)\times U(1)\times U(1)^3$.  
The gauge field of the extra $U(1)$'s can mix with the gauge field of the  $U(1)_Y$ 
in the kinetic term which leads to interactions between the visible sector and the hidden sector.
 
 As mentioned above 
 the mixing between the extra $U(1)$'s and $U(1)_Y$  can occur via kinetic mixing and via 
 Stueckelberg mass mixing.
 Of course, one may have both, i.e.,  kinetic mixing as well as  Stueckelberg mass mixing.
 The mixing involving the kinetic mixing vs the Stueckelberg mass mixing lead to rather different 
 types of interactions as explained below. 
 Thus for the case when  there is only kinetic mixing of the $U(1)$ gauge field of the visible and that of the 
hidden sector,  the visible sector photon  couples only to the visible sector matter 
while the hidden sector photon couples to both the
 hidden sector matter as well as the visible sector matter. This can be seen explicitly from Eq.(\ref{kst}).
 When one has both kinetic mixing as well as Stueckelberg mass mixing among the $U(1)$ gauge fields
  the photon of the visible sector also couples to matter in the hidden sector and this coupling is milli weak.
  For the case of no kinetic mixing and only Stueckelberg mixing the photon of the visible sector also 
  couples to the hidden sector matter. So in all three cases one finds that there is mixing between the
  visible sector and the hidden sector.  These mixings can give rise to observable effects. 
  Thus, for example, the hidden sector photon or $Z'$ particle 
  can decay into visible matter producing
  a very sharp resonance ~\cite{stueckelberg}
  (for a review of $Z'$ physics  more generally see~\cite{Langacker:2008yv}).   
   In the supersymmetric extension of the Stueckelberg mechanism 
  one has visible sector neutralinos interacting with hidden sector neutralinos 
  producing an LSP which is extra weakly interacting. In addition to the $U(1)$ portals there can be 
  other portals such as the Higgs portal which can produce communication between the hidden
  sector and the visible sector!\cite{Patt:2006fw}.
 
%%%%%%%%%%%%%%%%%%%%%%%%%%%%%%%%%%%

\section{The cosmic frontier\label{sec4}}
The cosmic frontier presents an important avenue for test of fundamental physics, since  models of the 
universe are deeply connected with  particle physics. Some of the areas where this connection
is particularly deep relates to  dark matter, dark energy, inflation, big bang cosmology,
baryogengesis and  leptogenesis.  We discuss some of these below.
   \subsection{Dark matter} 
  While the nature of dark energy remains somewhat of a mystery,   
  there are a signifiant number of candidates
  for dark matter. One of the most widely discussed possibility is that of 
   weakly interacting massive particles (WIMPs).
 Thus supergravity models provide the neutralino and the sneutrino  as possible  WIMP candidates for dark matter.
The neutralino  as a possible candidate for dark matter  was proposed soon after the formulation of
supergravity grand unified models~\cite{Goldberg:1983nd} and  here it appears as the 
 LSP over most of the parameter space of models  \cite{Arnowitt:1992aq} and with R -parity, it becomes a candidate for dark matter.
An important constraint on model building is  the constraint on relic density given by  PLANCK~\cite{Ade:2015xua}.
This means that at the very least the relic density not exceed the PLANCK~\cite{Ade:2015xua}
  limit of $\Omega h^2 \sim 0.12$. 
 While this constraint is generally easy to satisfy for sfermion masses in a wide class of supergravity models, it  becomes more difficult 
 if the neutralino is  mostly a bino. In this case the neutralino annihilation cross section becomes small 
making the satisfaction of the 
 relic density more difficult.  The satisfaction of the relic density here requires co-annihilation ~\cite{Griest:1990kh,Bell:2013wua,Baker:2015qna}
 which implies that there be one or more sparticles (NLSP) lying in the vicinity of the LSP. 
  However, coannihilation implies a compressed spectrum which leads to soft final states in the decay of the   
    NLSPs making the detection of  supersymmetry more
  difficult. Such a situation occurs for supergravity  models with sfermion masses which lie in the 10-100 TeV  region which then leads to a
  compressed spectrum to satisfy the relic density constraint.  
 Regarding the direct detection of dark matter, 
  the sensitivity of the dark matter detectors has been increasing steadily  and recent experiments (XENON100, PandaX, LUX) have 
  reached a sensitivity in the range of $10^{-45}- 10^{-46}$ cm$^2$ in spin-independent WIMP-nucleon cross-section.  
  It is projected that  by the year 
  2030 the sensitivity of direct detection experiments will reach the neutrino background of $10^{-49}$ cm$^2$   
  (for a review see~\cite{Undagoitia:2015gya})
  and beyond that the direct detection  of dark matter would be
  more challenging as it would require more accurate simulations of the neutrino backgrounds.
   
  Dark matter may also be detected by indirect means. Thus a long lived dark matter particle may decay 
  producing remnants which may be detectable. This is the case, for example, for a heavy neutral lepton which 
  may decay into a neutrino and a photon where the photon would be detected. 
    Indirect detection of dark matter can occur via detection of anti-matter produced in the annihilation of two dark
    matter particles.  
   For instance two neutralinos may annihilate 
  into $f\bar f$ producing antiparticles which can be detected. Various satellite experiments ~\cite{satellite-exp}
are exploring the presence of dark matter by these indirect means (for a review see~\cite{Slatyer:2017sev}).
 Aside from the dark matter particles  mentioned above there are a whole host of dark matter candidates
  whose masses range from $10^{-22}$ eV to the Planck mass. These include ultra light dark matter, sub GeV 
  dark matter ~\cite{Battaglieri:2017aum,Crisler:2018gci}, PeV  scale dark matter~\cite{Blanco:2017sbc},  
  extra-weakly interacting dark matter~\cite{Feldman:2006wd},  
    self-interacting dark matter, asymmetric dark matter~\cite{Kaplan:2009ag},   
    dynamical dark matter (see \cite{Dienes:2017zjq} and the references therein),  
  dark matter from extra dimensions (see~\cite{Rizzo:2018ntg} and the references there in)  
   and many others~\cite{ArkaniHamed:2008qn}.  
     Dark matter candidates whose production cross section is large enough to be produced at the LHC and other future 
   colliders could be detected as missing energy~\cite{LHC-dark}.    
 Below we discuss the possibility of an ultralight boson as a dark matter candidate.

  \subsection{An ultralight boson as dark matter in supersymmetry, supergravity and strings \label{sec4.2}} 
 Recently it has been pointed out that while the
 $\Lambda$CDM model works well for cosmology at large scales,  there are  issues at scales less than 
  $\sim10$kpc ~\cite{Weinberg:2013aya}.
Specifically one of the issue is the so called 
Cusp-Core problem. Here  N-body simulations show that CDM collapse leads to cuspy dark matter (NFW profile) halos.
 However, the observed galaxy rotation curves are better fit by constant dark matter density cores (Berkert profile).
The second issue concerns the so-called ``missing satellite'' problem. Here the CDM predicts too many dwarf galaxies.
Both of these problems could be solved by  complex dynamics and baryonic physics with WIMPs~\cite{Governato:2012fa}.
Another possibility mentioned to take account of cosmology at small scales, i..e, at scales less that $10$kpc is the ultralight dark matter
with a wavelength order $\sim 1$ kpc. One candidate is an ultralight boson~\cite{ultralight}
with mass $\mathcal{O}(10^{-22})$\,eV such as  an axion which is not 
 a QCD axion~\cite{bcd-axion} 
 but likely a string axion~\cite{Svrcek:2006yi}) with an  axion decay constraint in the range  $10^{16}\ {\rm GeV}\leq f \leq 10^{18}\, \GeV$.
 We note that this  range for the decay constant is  much larger  than  for the   
 QCD  case where  $10^9 ~{\rm GeV} \lesssim f  \lesssim 10^{12} \GeV$.
 Here
  the lower bound comes from red giant cooling 
and the upper bound comes from  consistency with the relic density constraint set 
by the misalignment mechanism. The string axion has greater flexibility in the range of allowed axion decay constant since it is not necessarily 
related to gauge dynamics of QCD axions. Such axions are rather ubiquitous in string theory. It is possible to construct models which 
lead naturally to an axion mass of size $\sim 10^{-21}$ eV  which can be found in~\cite{ultralight}.

\subsection{Dark energy}
The necessity of dark energy arises due to the observed accelerated expansion of the universe. In FRW universe the 
acceleration is governed by the Friedman equation 

\begin{align}
\frac{\ddot R}{R}= - \frac{4\pi G}{ 3 \pi} ( \rho +3 p)\,,
\end{align}
where $R(t)$ is the scale factor,  $\rho$ is the matter density and $p$ is the pressure. For accelerated expansion one needs
$(\rho + 3p) < 0$ which implies that  $w$ defined by $w= p/\rho$  obeys the constraint
\begin{align}
 -1\leq w < -\frac{1}{3}\,.
\end{align} 
Here the case $w =-1$ corresponds to  acceleration due to a cosmological constant. The current data from the PLANCK 
experiment~\cite{Ade:2015xua} indicates
\begin{align}
w = -1.006 \pm 0.045\,.
\end{align}
A value of $w$ greater than $-1$ would indicate the existence of something like quintessence (for a review see~\cite{Tsujikawa:2013fta})
and if $w<-1$ we will have a new sector involving
 something like phantom energy~\cite{Caldwell:1999ew}.
%~\cite{Caldwell:2003vq}

%it would indicate vacuum instability requiring 
% an analysis to guarantee the instability to be consistent with the current age of the universe.

\subsection{Inflation, spectral indices, non-Gaussianity}

As is well known inflationary models resolve a number of problems associated with Big Bang cosmology, i.e.,   
the flatness problem, the horizon problem,
and the monopole problem~\cite{early-inf}.
In inflation models  
quantum fluctuations at the time of horizon exit  contain  significant information on the 
 characteristics of the inflationary model~\cite{Mukhanov+}. 
 The cosmic microwave background (CMB) radiation anisotropy  encodes such information and its extraction 
 can help  discriminate  among   models. Analyses using  data from the PLANCK experiment ~\cite{planck}
   have been  helpful in putting significant constraints on models of  inflation eliminating some  and 
 constraining others. 
  Successful inflation requires pivot number of 
  e-foldings in the range  $N_{\rm pivot}=[50,60]$.   Experimentally measurable quantities are the ratio $r$ of the power spectrum for tensor 
 perturbations and the power spectrum of scalar perturbations, and the spectral indices of scalar perturbations $n_s$ and of tensor perturbations $n_t$
 (see section \ref{sec9.2}).
 The Planck experiment gives the following constraints on $r$ and $n_s$
\begin{align}
 r & <0.07\, (95\% {\rm CL})\,,\non
 n_s& = 0.9645 \pm 0.0049\, (68\% {\rm CL})\,,
\label{data}
\end{align}
while $n_t$ is currently unconstrained. As  noted above  these results have put significant constraints  on models of inflation.

  An interesting class of models are the so called natural models of inflation which involve an 
 axionic field~\cite{natural-inf}.
  The simplest version of this model based on the QCD axion 
 is currently highly constrained. 
 Thus the natural inflation model involves an axion potential of  the form
 \begin{align}
V(a) = \Lambda^4 \left(1+ \cos(\frac{a}{f})\right)\,.
\end{align}
Here  $f$ is the axion decay constant and as mentioned in section \ref{sec4.2} 
is constrained so that $ 10^9 ~{\rm GeV} \leq f \leq 10^{12} ~{\rm GeV}$. 
However, in natural inflation one  requires  $f> 10 M_P$. One procedure to overcome it is the so called alignment 
mechanism~\cite{Kim:2004rp}.  
 The problem of the axion decay constant can be 
more easily overcome in an axion landscape ~\cite{Nath:2017ihp}. Here  one can generate an axion 
 potential  dynamically  with stabilized saxions and with  breaking of the shift symmetry by non-perturbative terms~\cite{Nath:2017ihp}.
Here 
successful inflation occurs with  the  desired pivot number of e-foldings in the range $N_{\rm pivot}=[50,60]$. 
Further, the ratio $r$ of the tensor to the scalar power spectrum and the spectral indices  
are found consistent with PLANCK data Eq. (\ref{data}).
It  is of interest to note that 
for slow roll  the ratio $r$ and the spectral indices are related by the constraint 
\begin{align}
n_s= 1-6\epsilon + 2 \eta, ~~~n_t = -2 \epsilon, ~~ r = 16 \epsilon\,,
\label{ns-nt-r}
\end{align}
where $\epsilon$ and $\eta$ are defined in section \ref{sec9.2}. 
Eq. (\ref{ns-nt-r})  gives 
 the standard single-field inflation result and the prediction $n_t=-r/8$. Effective single field models
reduced from multi-field inflation can generate small deviations from this prediction which can in principle be
tested by more accurate determination of $r$ and of  the spectral indices~\cite{Nath:2017ihp}.

Another experimentally testable quantity that can discriminate among inflation models is non-Gaussianity $f_{\rm NL}$ (see section \ref{sec9.3}).
Roughly speaking  $f_{\rm NL}$  depends on the sound speed~\cite{non-gauss}
\begin{align}
f_{\rm NL}\simeq c (\frac{1}{c_s} -1)\,,
\end{align}
where $c=O(1)$ and $c_s$ is the sound speed defined by
\begin{align}
c^2_s= \frac{p_{,\beta}}{\rho_{,\beta}}\,.
\end{align}
Here $p$ is the pressure and $\rho$ is the density in the early universe at the time of inflation, and $\beta= \dot \phi^2$, where $\phi$ is the inflaton field. For models with canonical kinetic energy $c_s=1$ and so the non-Gaussianity vanishes. But non-vanishing non-Gaussianity can
arise in a more general class of models with non-canonical kinetic terms 
such as the Dirac-Born-Infeld models. The analysis of ~\cite{nongauss-test}
  indicates that non-Gaussianity can be tested in the near   future experiments 
provided $|f_{NL}|> 5$.

 The simplest example of a Dirac-Born-Infeld (DBI) lagrangian for a scalar field  is  given by
 \begin{align}
 \mathcal{L}_{DBI}= -\frac{1}{f} \sqrt{ 1+ f \partial_\mu\phi  \partial^\mu\phi} - V(\phi).
 \end{align}
  Here one finds $c_s$ 
  \begin{align}
  c_s= \sqrt{1- f \dot \phi^2}\,.
  \end{align}
 For small $c_s$ one needs $f\dot \phi^2$ to be close to $1$.
 However, at the same time for successful inflation one needs the  pivot number of e-foldings in the range  [50-60] and
 also one needs to satisfy the constraints from the PLANCK experiment~\cite{Ade:2015xua}.  This is non-trivial. 
    We discuss now the extension of the DBI lagrangian to the case of supersymmetry. The simplest example 
    of this is given in \cite{susy-dbi} and discussed in section \ref{sec9.6} and a more realistic model is discussed 
    in \cite{Nath:2018xxe}
.  In  \cite{Nath:2018xxe}
 an analysis is carried out for an axionic DBI inflation model 
     with a $U(1)$ global symmetry which is broken by instanton effects. Here  the parameter
     space of the model was analyzed so that one has the desired pivot number of e-foldings and also one satisfies
     the PLANCK experimental constraints.  In this case one finds that while the model leads to a non-vanishing
     non-Gaussianity, the predicted value of non-Gaussianity is significantly smaller than the projected lower bound of 
     $|f_{NL}| >5$ in the improved near future experiments.

\section{Unification frontier\label{sec5}}
Over the past decades progress has occurred on  unification
on two fronts:  grand unified theories and  strings.
Grand unified theories are effective theories  expected to be valid at scales 
 $E \leq M_{G}< M_{Pl} $ where $M_G$ is the grand unification scale. Above this scale one is in the regime of quantum gravity which
 is properly described in the framework of strings. 
One of the strong constraints in searching for acceptable grand unified groups is that
at the very least they must possess  chiral representations in which matter can reside.
The only possibilities here are  $SU(N), SO(4N+2), N\geq 1, E_6$.
Among these the leading candidate currently is the group $SO(10)$.
A nice property of this group is that one full generation of quarks and leptons can be
accommodated in the 16-plet spinor representation of $SO(10)$.
However, the not so nice feature of $SO(10)$ unification is that 
there are many possibilities for the Higgs representations that break the group $SO(10)$ down to the standard model gauge 
group and then to the residual gauge group $SU(3)_C \times U(1)_{\gamma}$. 
Thus one needs at least three
Higgs representations to break the symmetry down to the residual gauge group $SU(3)_C\times U(1)_{\gamma}$,
i.e., a $16+\overline{16}$ or $126+\overline{126}$ to reduce the rank, a $45, 54$ or $ 210$ to 
reduce the group to the SM gauge group and then $10$-plets to break the symmetry down to 
$SU(3)_C\times U(1)_{\gamma}$. This problem could be overcome   by use of a unified Higgs sector  where one
uses  $144+\overline{144}$~\cite{Babu:2005gx}. 
Another issue concerns the doublet- triplet splitting. In $SO(10)$ this problem could be
overcome by use of $560+\overline{560}$~\cite{Babu:2011tw} which has a unified Higgs sector producing only 
one light Higgs doublet and makes all the Higgs triplets heavy and 
 is an extension of the missing partner mechanism \cite{Masiero:1982fe} .

While  GUTs  are  consistent effective theories they are not UV complete in themselves but could either be remnants of 
string theory or belong to what one may call 
``swampland''~\cite{ArkaniHamed:2006dz}.
 The probability of a theory in the swampland being a theory of quantum gravity is deemed very small.
  Currently the only theories of quantum gravity are the five  well known strings, i.e., 
  Type I, Type IIA, Type IIB, heterotic $SO(32)$,  and heterotic $E_8\times E_8$.
The Type I and Type II strings contain $D$ branes. The D branes
can support gauge groups and chiral matter can exist at the intersection of D branes. 
The  various string types can arise from the so called M-theory whose low energy limit is 11D supergravity.  
Most of the model building in string theory has occurred  in the following areas: 
 heterotic
$E_8\times E_8$/Horava -Witten theory  where Horava-Witten theory~\cite{Horava:1995qa}
 is the strong coupling limit of $E_8\times E_8$,
and  Type IIB/F-theory, where F theory~\cite{Vafa:1996xn}  is the strong coupling limit of Type IIB strings.
 One can also compactify M-theory directly such as on a manifold $X$ of $G_2$ holonomy~\cite{G2}.
One  problem in working with string models is that they possess a huge number of vacuum states.
i.e., as many as $10^{500}$ for type IIB~\cite{Douglas:2006es},  and in the context of $F$ theory as many as $10^{272000}$~\cite{Taylor:2015xtz}
  and thus the search for the right vacuum state that describes our universe is a daunting task.
  
At the phenomenological level there are  constraints on string models even more stringent than  those on GUTs. Thus string models
in four dimensions have the problem of the so-called  moduli stabilization.  Moduli are massless fields which have gravitational
interactions with the visible sector fields. They need to be stabilized to extract meaningful physics from the theory.
 Progress has occurred in this direction in
 the past decade and a half through the work of KKLT~\cite{Kachru:2003aw} 
 and in the frame work of the large volume scenario~\cite{Balasubramanian:2005zx}.
	Another instance where string models are more constrained than GUTs relates to the unification of 
	the dimensionless parameter $\alpha_{\rm GR} = G_N E^2$, where $G_N$ is Newton's constant, 
	with the 
 three gauge couplings of the standard model (for a review see~\cite{Dienes:1996du}). Using the normal spectrum of the MSSM one finds
 that the gauge couplings do not unify with the gravity coupling $\alpha_{\rm GR}$.
  Several possibilities exist for resolving this issue.  One possibility relates to a new dimension 
 opening up below the scale $M_G$ under the assumption that matter resides on the four dimensional 
 wall while  gravity propagates in the bulk  which speeds up  the evolution of the gravity coupling  
 allowing for a unification of gauge and gravity couplings and this can occur much below the Planck scale. 
 A further refinement of this ides is a much lower  string scale~\cite{low-string-scale}.

 \subsection{Tests of Unification}
 Since unification of particles and forces is the central theme of high energy physics, it is imperative to ask what are the tests of such a unification.
One of these is the well known unification of gauge couplings~\cite{Georgi:1974yf} which works well for supersymmetry with the
sparticle masses at the electroweak scale~\cite{Dimopoulos:1981yj}.
 It is reasonable to ask what happens if the sfermion masses are much higher and lie in the 10-100 TeV range
 which is a possibility in view of the high mass of the Higgs boson in supersymmetry. 
 An analysis shows that the unification works equally well and sometimes even better with squark masses
 in the 10-100 TeV range~\cite{Aboubrahim:2017wjl}.  
Further as is well known unified models lead to instability for the proton (for reviews see  \cite{Nath:2006ut,Raby:2008pd,Hewett:2012ns}).
 One common prediction in non-supersymmetric and supersymmetric  GUTS as well in strings is the existence of the  proton  decay mode
 $p\to e^+ \pi^0$ which is generated by the exchange of lepto-quarks and a rough estimate of it gives  the decay width
  $\Gamma (p \rightarrow e^+ \pi^0) \simeq  \alpha_G^2 \frac{m_p^5}{M_V^4}$ where $\alpha_G=g_G^2/4\pi$
 with $g_G$ being the unified coupling constant, and $M_V$ the lepto-quark mass. It leads to a partial lifetime
 of  $\tau (p \rightarrow e^+ \pi^0) \simeq 10^{36\pm1} yrs$.   
   This mode may allow one to discriminate between GUTs and strings. Thus generic D brane models allow only $10^2\overline{10}^2$ SU(5) type dimension six operators which give $p\to \pi^0 e^+_L$ while SU(5) GUTs also have $10\bar{10}5\bar 5$ which allow $p\to \pi^0e^+_R$  ~\cite{Klebanov:2003my}.
Additionally the $SU(5)$ GUT models  also give the decay $p\to \pi^+\nu$ which is not allowed in generic D brane models.  However, there exist special regions of intersecting D brane models where the operator $10\overline{10}5\bar 5$ is indeed allowed and the 
 purely stringy proton decay rate can be of the same order as the one from SU(5) GUTs  including the mode $p\to \pi^+\nu$ ~\cite{Cvetic:2006iz}. 
 The current experimental limit from    Superkamiokande is the following: 
 $\tau(p\to e^+\pi^0) >2\times 10^{34} ~{\rm yrs}$~\cite{TheSuper-Kamiokande:2017tit}.  
  More sensitive proton decay experiments are needed to push the experimental limits.
  Thus the proposed  experiment   Hyper-K is expected to achieve a  sensitivity of
   $\tau(p\to e^+\pi^0) > 1\times 10^{35}~{\rm yrs}$~\cite{Abe:2011ts}. 

Observation of the strange decay mode of the proton, i.e.,
 $p\to \bar \nu K^+$ would be a remarkable evidence for both grand unification and supersymmetry/supergravity.
 This is so because this mode is suppressed in non-supersymmetric grand unification and is dominant in supersymmtry/supergravity
 unification.  The current experimental bound on this decay 
 from Super-Kamiokande  is $\tau(p\to \bar \nu K^+) > 4\times 10^{33} {\rm yrs}$.   Larger  sensitivities are expected  in the future at the 
proposed Hyper-K  experiment where sensitives of $\tau(p\to \bar \nu K^+) > 2\times 10^{34} ~{\rm yrs}$
could be achieved~\cite{Abe:2011ts}. Another remarkable aspect of proton decay is that it can in principle distinguish between 
 the grand unification groups $SU(5)$ and $SO(10)$. Thus the group $SO(10)$ allows for $|\Delta (B-L)|=2$ interactions while 
 the group $SU(5)$ does not. Such interactions can lead to proton decay modes $p\to \nu \pi^+, n\to e^- \pi^+, e^- K^+$
 which would not be allowed in $SU(5)$ and would indicate $SO(10)$ or a higher rank group as the underlying gauge group. Further these
 interactions also allow for baryogenesis,  and $n-\bar n$  oscillations (For a review of the literature and some recent works
see ~\cite{B-L-violating}).
  
 As discussed in section \ref{sec2} a large value of the weak supersymmetry scale  in the TeV region
 is natural in radiative breaking if it lies 
 on the hyperbolic branch.  Further, it generates a desirable loop correction needed to lift the Higgs boson mass above
 the tree level, and scalar masses in the tens of TeV lead to a unification of gauge couplings consistent with experiment.
 Further, it helps stabilize the proton from interactions arising from the baryon and lepton number violating dimension five operators.
  As mentioned in section \ref{sec2}
  an additional benefit of a large value of the scale of weak scale supersymmetry is that 
 it solves the so called gravitino problem. 
Thus from the early days of  supergravity models it is known that   stable gravitinos produced in the early universe can 
overclose the universe if the gravitino
mass exceeds 1 keV~\cite{Pagels:1981ke}.   Unstable gravitinos  decay with gravitational strength  and their 
 decay modes are { $\tilde{G}\rightarrow \tilde{g}g, ~  \tilde{\chi}_1^{\pm}W^{\mp}, ~\tilde{\chi}_1^0\gamma,
~\tilde{\chi}_1^0 Z$}. These decays could upset the BBN if they occur  after the BBN time, i.e., $(1-10^2)$s. 
It turns out  that $10-100$ TeV gravitinos decay before BBN. One of the end products of the gravitino decay 
is  the neutralino which contributes to the relic density. Thus the 
neutralino density has two components in this case: 
 $\Omega_{\tilde{\chi}^0_1} =  \Omega^{\rm th}_{\tilde{\chi}^0_1} +  \Omega^{\tilde G}_{\tilde{\chi}^0_1}$.
 One part ($\Omega^{\rm th}_{\tilde{\chi}^0_1}$) 
 is thermal and the other  ($\Omega^{\tilde G}_{\tilde{\chi}^0_1}$) is 
  non-thermal. The decaying gravitinos produce a non-thermal contribution to the relic density. 
However,  for reheat temperatures up to $10^{10}$ GeV, the non-thermal contribution to the relic 
density is negligible~\cite{Aboubrahim:2017wjl}.\\

 \subsubsection{Can all the string vacua be tested?}
   As mentioned in the beginning of this section, 
while strings  provide a framework for quantum gravity,  the number of allowed vacua is large;
around $10^{500}$ for type IIB string and even larger for  F-theory.  One might speculate that in the future 
fast computers could  check them all. 
 To see if this is realistic we note that the age of the universe $\tau_{\rm uni}$ since the Big Bang is   $\sim4.32 \times 10^{17}$ s. 
   Suppose we make the (inspired) estimate that in future fast computers may check as many as $10^{10}$ vacua/s, which then
   implies that it would take $\sim 10^{472} \tau_{\rm uni}$ to check all. 
    While machine learning tools could help such techniques by definition are not exhaustive.     
    However, some avenues for finding the vacuum state that describes our universe are still possible. 
    For example, such a discovery could be purely accidental, i.e.,  one stumbles on the
    right vacuum by pure luck although the chances for this to happen are remote. Another possibility is that standard like vacua within strings
    constitute a non-negligible fraction of the full set and thus dedicated searches which even explore a minuscule fraction of the 
    string vacua will find standard model like vacua.  
           We note, however, that  some generic features of strings could  be tested
    even without having a concrete string model. Such tests could involve phenomena on the interface of particle physics 
   and cosmology, or relate to predictions such as appearance of resonances for low scale strings see, e.g., \cite{collider-extra-dim,Anchordoqui:2015uea}
or find relation between string scale physics that are generic to strings 
and low energy phenomena (see, e.g., ~\cite{ultralight,G2,Klebanov:2003my,Cvetic:2006iz,Nath:2002nb,Raby:2011jt,Dhuria:2014fba}).

\section{Challenges for High Energy Physics and Cosmology \label{sec6}}

We discuss now some of the  challenges for high energy physics and for cosmology over the coming years.
 On the experimental side one of the challenges is to establish the 
 nature of dark matter, i.e., whether it is a  WIMP, an axion, fuzzy dark matter or 
 something else and determine its properties, i.e., mass, spin and interactions  with the standard model fields. 
    The importance of 
 discovering sparticles which are central  to establishing supersymmetry as a fundamental 
  symmetry of nature cannot be overemphasized. 
  A related challenge is the discovery of additional Higgs bosons which arise in the most popular
  models of supersymmetry.
On the unification frontier, the discovery of proton decay, which exists in all unified models,
and the specifics of its branching ratios will  shed light on the nature of the underlying model. 
Closely connected with proton decay is determining the  nature of neutrino mass hierarchy, i.e., whether it is a normal hierarchy or an
 inverted one, and the more challenging task of fixing the absolute scale of the neutrino masses.
  Aside from the above there are a variety of other experimental challenges which would shed light on the 
 nature of fundamental interactions. Thus  evidence for a non-vanishing edm for leptons and quarks  
 would indicate existence of new sources of CP violation beyond what one has in the standard model. 
 Similarly identification of flavor violating processes in the Higgs sector will be a guide to constraining 
 particle physics models.  One of the hall marks for a variety of grand unified models is  the presence of vector like 
 particles. 
  While most of these vectorlike states 
    are high mass, some of them could be low lying and discoverable.
 The detection of
such states will point to the possible matter  and Higgs representation that appear in a GUT model or  in a string model.

 Experimental evidence for the existence of a hidden sector is of great interest and  a challenge 
  for experiment.  As discussed earlier, hidden sectors arise in supergravity models, in string 
  and D brane models and they can communicate with the visible sectors via Planck suppressed
  operators, via kinetic and \st mixing between $U(1)_Y$ and the $U(1)$ of the hidden sector,
  and via Higgs portals.  Experimentally, the presence of the hidden sectors
  could manifest in a variety of ways, such as via very sharp $Z'$ resonances, via extra weakly interacting
  dark matter and via hidden sector photons interacting with matter in 
  the visible sector.  Observations of such effects would
provide evidence of a new  sector of physics beyond the standard model.

 On the cosmology side improved  determination of the tensor to scalar power spectrum ratio $r$ and of the spectral indices 
$n_s$  and $n_t$ are central to fixing more narrowly the allowed inflation models. For slow roll one has $n_t+ r/8\sim 0$
and any significant deviations from it will have implications for the underlying particle physics model of the inflaton.   
Further, the experimental constraints on non-Gaussianity in improved experiment in the future will put more stringent
 limits  on inflation models. 
On the theoretical side the major challenge is finding the framework where both the particle physics models  and  the inflation 
models have a common origin. 
An outstanding challenge for  particle physics and for cosmology relates to deciphering the nature of dark energy which  contributes 
a major portion of mass to the universe, i.e., $\Omega_\Lambda =0.72$.  
Currently the data fits well with the $\Lambda$CDM model. However, the smallness of the cosmological constant $\Lambda$ 
is puzzling  (for a review see~\cite{Weinberg:1988cp}).
 Here a  more accurate determination of the equation of state $p=w\rho$, i.e., of $w$ would be great significance.
   The current experimental determination gives
  $ w = -1.006 \pm 0.045$~\cite{Ade:2015xua}. 
  An appreciable deviation from $-1$ would indicate
 new physics beyond $\Lambda$CDM. Thus $w>-1$ might indicate something like quintessence and $w<-1$ 
 would open up a new  sector of physics~\cite{Carroll:2003st}. 
% would leads to vacuum instability and would  need to be investigated carefully to make sure that the timescale  for such instability
% is greater than the age of the universe~\cite{Carroll:2003st}.

\section{Future high energy colliders\label{sec7}}
LHC2 will run at 13 TeV center of mass energy till the end of 2018 and by that time CMS and ATLAS are each expected to 
collect about 
150 fb$^{-1}$ of integrated luminosity. LHC will then shut down for two years in the period 2019-2020 for 
upgrade to LHC3 which will operate at 14 TeV in the period 2021-2023. In this period each of the detectors will additionally 
collect up to 300 fb$^{-1}$ of data. LHC will then shutdown for a major upgrade to high luminosity LHC (HL-LHC or LHC4) for a two
and a half years in the period 2023-2026 and will resume operations in late 2026  and run for an expected 10 year period
till 2036. In this period it is expected that each detector will collect  about 3000 fb$^{-1}$ of additional data.

There are two main types of colliders being thought of for the future. One type are $e^+e^-$  colliders 
beyond the LEP II energy and  the other are $pp$ colliders beyond the LHC energy of 
14 TeV. Thus three different proposals for   $e^+e^-$ colliders  have been discussed. These include the   
International Linear Collider (ILC)  in Japan,  the Circular Electron-Positron Collider (CEPC) in China
 \footnote{\small C. N. Yang has argued against the construction of the Circular Electron-Positron Collider (CEPC)
  in China listing a variety of reasons which include scientific, financial and societal.  A response to these
  is given by David Gross~\cite{Gross:2016ild}.} and the
 Future  Circular  Collider (FCC) at CERN.
 The $e^+e^-$ collider will essentially be a Higgs boson factory. The prospects for an $e^+c^-$ collider 
 have been enhanced because of the discovery of the Higgs boson. Currently the branching ratios of the
 Higgs boson are consistent with the standard model but the error bars are significant. An $e^+c^-$
 machine will provide much more accurate measurement of the branching ratios and a deviation from 
 the standard model prediction will be an indication of new physics beyond the standard model.
 
 The $e^+e^-$ collider  will  likely operate at an energy in the vicinity of  $240$ GeV  to optimize the 
 cross section for $e^+e^-\to ZH$ production sometimes referred to as  Higgsstrahlung, as it is a process
 better suited for studying the properties of the Higgs boson than using the $2H$ final state. The reason for that
 is because  the leptonic decays of the $Z\to \ell^+\ell^-$   
 provide a cleaner signature for  the Higgs boson final state 
 than the  production of the two Higgs state which will generate a negligible cross section for the decay into
 electrons or muons.  As noted 
  the $e^+e^-$ machine  is primarily a Higgs boson factory and not the 
 prime machine for the discovery of sparticles. For that reason 
 proton-proton colliders remain the main 
 instrument for the discovery of supersymmetry.

\begin{figure}[H]
 \centering
   \includegraphics[width=0.36\textwidth]{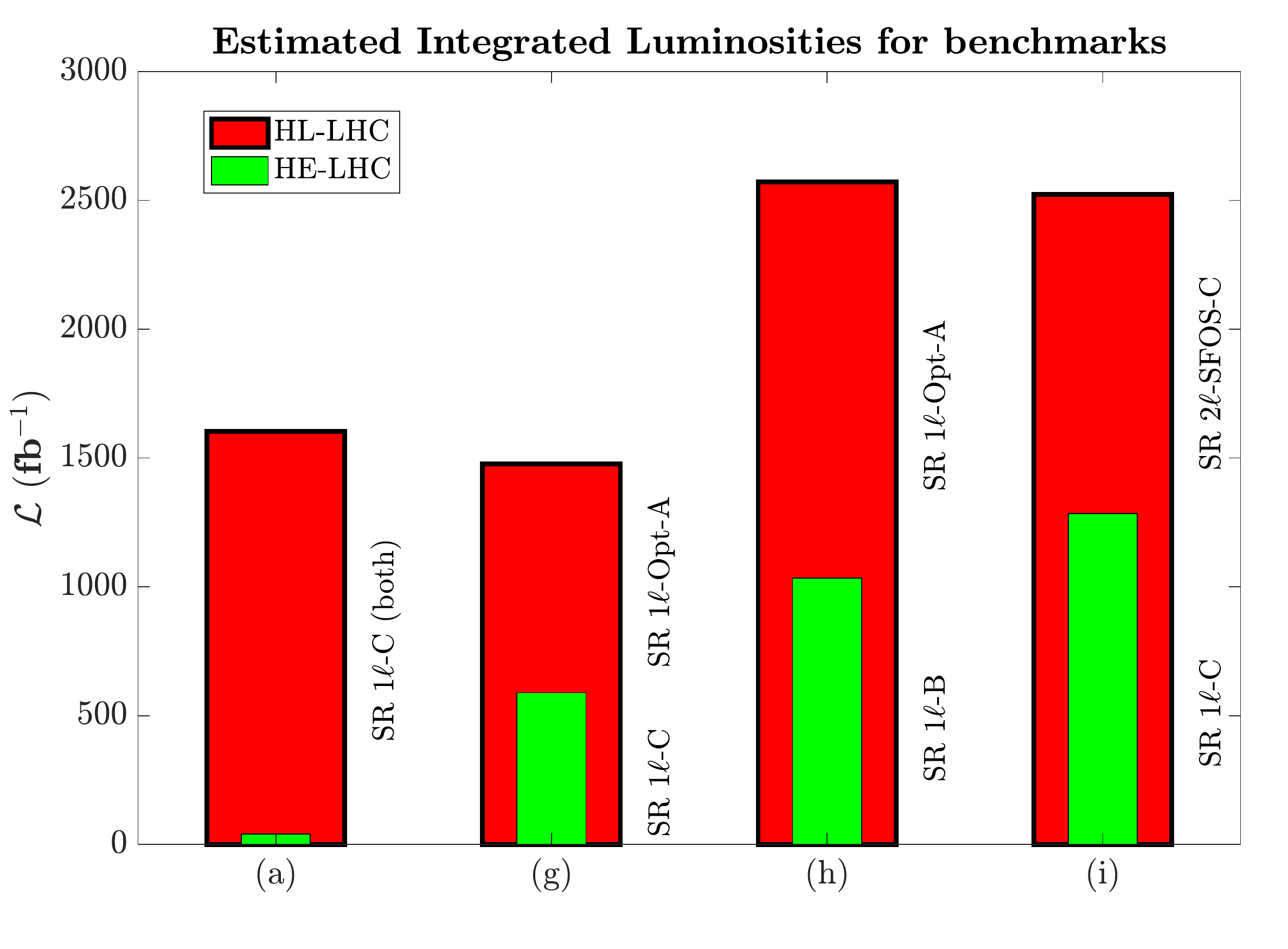}
      \includegraphics[width=0.36\textwidth]{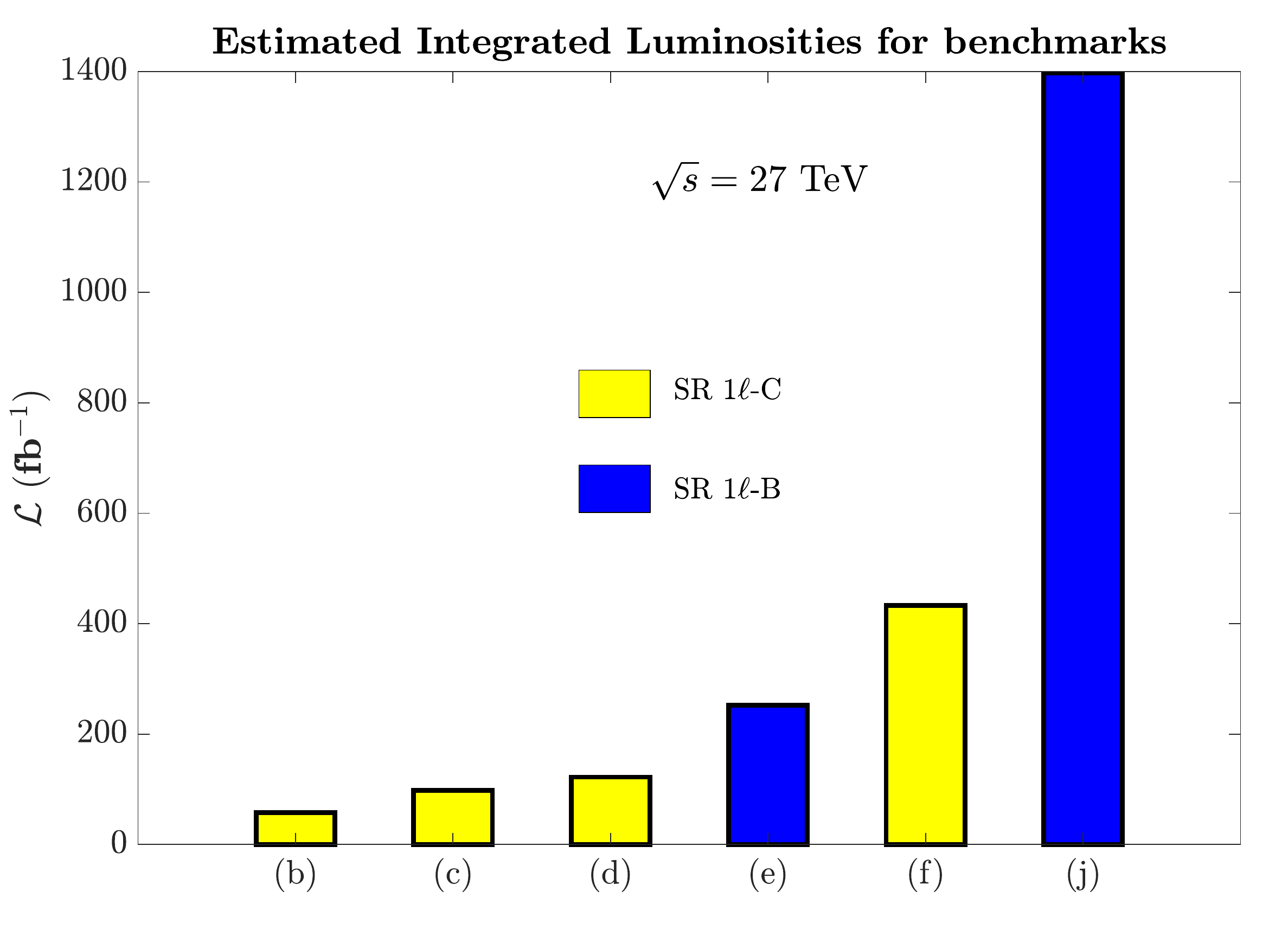}   
    \caption{
    Left panel:
      A comparison of the discovery potential of HL-LHC at 14 TeV and HE-LHC at 27 TeV 
 using    estimated integrated luminosities, $\mathcal{L}$, for a 5$\sigma$ discovery
of the benchmark points points (a), (g), (h) and (i) whose input parameters are as given in~\cite{Aboubrahim:2018bil}.
    Right panel: HE-LHC analysis of model points which are not accessible at HL-LHC but would be accessible at HE-LHC where
    the input parameters  of models (b)-(f) and (j) are as given in ~\cite{Aboubrahim:2018bil}. Taken from~\cite{Aboubrahim:2018bil}. }
    	\label{lum1-lum2}
\end{figure}

It is very likely that the LHC after the upgrade and with its optimal  integrated luminosity will discover one or more 
 low lying sparticles. However, irrespective of whether or not the sparticles  are discovered at the LHC at $\sqrt s=14$ TeV, 
one would need  a larger energy  machine to discover sparticles up the ladder and for those sparticles  that are discovered 
at $\sqrt s=14$ TeV, the larger energy machine would be helpful in studying their properties more accurately.
Two  types of higher energy pp colliders have recently been discussed. One of these is a  100 TeV machine (for physics opportunities at a 100 machine see,
e.g., \cite{Arkani-Hamed:2015vfh,Golling:2016gvc}).
Thus one proposal  being considered by  the Future Circular Collider (FCC) study at CERN is  a 100 TeV collider to be installed in a 100 km tunnel in the Lake Geneva basin. Also a 100 TeV Super proton-proton Collider (SppC) is being considered
in China. More recently a third possibility, i.e., a  high energy LHC (HE-LHC),   under consideration by the FCC study at CERN, would use the existing tunnel at CERN 
with FCC technology magnets to achieve a center-of-mass energy of $\sqrt s=27$ TeV and at the significantly enhance luminosity of 
$2.5\times 10^{35}$ cm$^{-2}$ s$^{-1}$.  An analysis of the discovery potential of this machine for the discovery of sparticles 
has been carried out in~\cite{Aboubrahim:2018bil}
 which analyzed a set of model points of supergravity unified models. In this analysis it is found
that the model points which are discoverable at the HL-LHC with a 
run between 5-8 years are discoverable  in a period of few weeks to $\sim 1.5$ yr at HE-LHC running at its optimal luminosity of $2.5 \times 10^{35}$ cm$^{-2}$s$^{-1}$.
Additionally HE-LHC can probe the parameter space beyond the reach of HL-LHC. Fig. (\ref{lum1-lum2}) gives a comparison of the discovery potential of 
HE-LHC vs HL-LHC.

\section{Conclusion\label{sec8}}
  One of the compelling reasons for the discovery of supersymmetry is the suppression of the quadratic divergence in the Higgs
  loop arising from cancellation of the quark loop contribution with the contribution of  
  squark loop.  One may compare this cancellation with the 
   GIM mechanism for the cancellation among loops involving up quark and the charm quark in the 
  flavor changing neutral current contribution to  the process $K^0 \to \mu^+\mu^-$.  Here the 
  cancellation is  up to one part  to 1 part in $10^{9}$ since
  $Br(K^0\to \mu^+\mu^-)/Br(K^+\to \mu^+ \nu_\mu) =  (6.84\pm 0.11) \times 10^{-9}$. 
  The GIM mechanism provided a strong motivation for the search for the charm quark which was eventually discovered. 
For the case of the Higgs boson  the cancellation is up to one part in $10^{28}$ to produce a Higgs boson mass
compatible 
with experiment. 
Further, as discussed in section \ref{sec2} the case for supersymmetry is stronger post
  Higgs boson discovery. 
    Currently there is no good alternative to \sugra as a workable paradigm that can allow us to extrapolate physics 
    from the electroweak scale to the grand unification scale.  Of course, a direct confirmation of supersymmetry would
    require discovery of one or more sparticles. 
    The most likely candidates here for discovery are the
  $ \tilde \chi^0,  \tilde \chi^{\pm}, \tilde g, \tilde t_1, \tilde \tau_1.$.
   There is a good chance that with 
 the full capability of the LHC ($\cl =3000$ fb$^{-1}$, $\sqrt s=14$ TeV)   one discovers supersymmetry.
At the same
 time a good idea to look ahead and plan for colliders with larger energy, such as the  high energy upgrade of the LHC to 27 TeV, i.e., HE-LHC, or a 
 100 TeV pp super collider, either in an enlarged 100 km ring at CERN, and/or a 100 TeV pp collider (SppS) in China.
  The higher energy $pp$ colliders can expedite the discovery of new physics. We discussed
  here examples where the model points of supergravity models 
  that might take up to five years for discovery at the  HL-LHC can be discovered at the 
   HE-LHC within few weeks. Further, HE-LHC can explore the parameter space beyond the reach of HL-LHC.  A similar situation 
   is expected for the higher mass Higgs bosons in supersymmetric models.  It is also pointed out that
   the discovery of supersymmetry associated with missing energy also constitutes discovery of supersymmetric dark matter.
   This is of importance since the direct detection experiments in the not too distant future will be achieving a  sensitivity 
   which approaches the neutrino floor. In that case, it would make the direct detection of dark matter more difficult. However, LHC
   can act as a dark matter factory for the production and study of supersymmetric dark matter even when their observation in direct
   detection may be difficult.
   
    We discussed also some front line issues related to cosmology. Thus some of the main issues in cosmology relate 
    to  inflation and dark energy. While there exist many inflation models,  several have been constrained and 
    others essentially eliminated by the PLANCK data~\cite{Ade:2015xua} on the tensor to scalar ratio $r$ of the power spectrum. 
    In future more accurate data on $r$ as well as on scalar and tensor spectral indices and on non-Gaussianity 
       as one expects in stage4 CMB experiment~\cite{Abazajian:2016yjj,Schaan:2016ois}
       will hopefully help us produce a standard model of inflation.      
           A similar situation exists with regard to dark energy.
      Currently the $\Lambda$CDM model works very well, but the equation of state given by $w=p/\rho$ has a spread around
      $-1$. More accurate determinations of $w$ from CMB-S4, LSST and EUCLID
      \cite{Abazajian:2016yjj,Schaan:2016ois,Laureijs:2011gra,Rhodes:2017nxl}      
      will have very significant implications regarding our view of the universe .
      Thus $w>-1$ will indicate that the dark energy may be something like quintessence, while $w<-1$
      would indicate a new sector of physics involving something like phantom energy. 
       In any case particle physics will play a greater role in the coming years 
      linking high energy physics and cosmology more closely. This is desirable from the unification view point since
      a true unified model must describe both high energy phenomena as well as phenomena at the cosmological scales.

 %%%%%%%%%%%%%%%%%%%%%
\section*{Acknowledgments}
This research was supported in part  by  NSF Grant PHY-1620526.
The content of this paper is based in part on the  summary talk given at the 25th International Conference on Supersymmetry
and Unification of Fundamental Interactions, at TIFR, Mumbai, December 11-15, 2017,  and talks at PLANCK2017, Warsaw, May 22-27, 2017,
and at  PASCOS2018, Cleaveland, Ohio, June 3-8, 2018. Discussion with a number of colleagues related to some aspects of the
material discussed here
are acknowledged. They include Amin Aboubrahim,   David Gross, James Halverson,  Cody Long, Brent Nelson, Maksim Piskunov,
Andrew Spisak,  Tomasz Taylor and Darien Wood.

\section{Appendix: Notation and basics formulae \label{sec9}}

\subsection{Natural heavy scalars in radiative breaking\label{sec9.1}}
Radiative breaking of the electroweak symmetry  has  the form 
\begin{equation}
\mu^2 + \frac{1}{4}M_Z^2= C_1m_0^2+ C_2 A_0^2+ 	 C_3m^2_{1/2}+C_4 m_{1/2} A_0 +  \Delta \mu^2_{loop}\,,
\label{rewsb1}
	\end{equation}
	where $C_1-C_4$ are as defined in~\cite{hb},	
and  $\Delta \mu_{loop}^2$ is the loop correction.  
 To exhibit the underlying geometry we can choose a renormalization group scale $Q$ so that
 the loop term vanishes.  In this case we can write the radiative breaking equation Eq. (\ref{rewsb1})  in the form 
 \begin{align}
 s^2 &= \xi^i g_{ij} \xi_j, ~~i,j=1,2,3,
 \label{s2}
\end{align}
where  $s^2= \mu^2+ \frac{1}{2} M_Z^2$,  $\xi^i=(m_0, A_0, m_{1/2})$,
$g_{11}= C_1$, $g_{22}= A_0$, $g_{33}=C_3$,  $g_{12}=0=g_{13}$,  $g_{23}=g_{32} = \frac{1}{2}C_4$. 
We may further write the above relation so that 
\begin{align}
 s^2 &= m_0^2g_{11} + \Delta^2,
 \end{align}
 where $\Delta^2= \sum_{i,j=2,3} \xi^i g_{ij} \xi^j$. 
In RG evolution, typically $\Delta^2>0$. However, $g_{11}$ does not have a fixed sign
and can move from being positive to being negative depending on the region of the parameter space one is in, giving rise to two branches of radiative
breaking which 
correspond to $g_{11} >0$ (ellipsoidal branch: EB) and $g_{11} \leq 0$ (hyperbolic branch: HB). 
HB contains focal points, focal curves and focal surfaces. 
Thus we can classify radiative breaking of the electroweak symmetry so that 
\begin{align}
g_{11}&>0,~~{\text{Ellipsoidal Branch: EB}}\,,\non
g_{11}&=0, ~~{\text{Focal Point region of HB}}\,,\non
g_{11}&<0, ~~{\text{Focal Curves, Focal  Surface region of HB}}\,.
\label{9.114}
\end{align}
Eq.(\ref{s2}) exhibits the fact that the 
  constraint of electroweak symmetry breaking is a relation between the parameter $\mu^2$ which preserves supersymmetry, and
the parameters $m_0, A_0, m_{1/2}$
which break supersymmetry.  One finds that for a fixed value of $s$, one gets naturally heavy scalars as $m_0$ can get large for the case $g_{11}\leq 0$ which is 
the HB region of radiative breaking of the electroweak symmetry.

\subsection{Kinetic and Stueckelberg mass mixings \label{sec9.4}}
In this Appendix we give details of kinetic mixing ~\cite{holdom} and of Stueckelberg mass mixing ~\cite{stueckelberg}.
For simplicity we consider the case of 
two gauge fields $A_{1\mu}, A_{2\mu}$ corresponding to the gauge  groups $U(1)$ and $U(1)'$ and 
we choose  the following Lagrangian
$\mathcal{L} =\mathcal{L}_{\rm kin} +  \mathcal{L}_{\rm St-mass}+ \mathcal{L}_1$ where

\beqn \mathcal{L}_{\rm kin} &=&
    - \frac{1}{4}F_{1\mu\nu}F_1^{\mu\nu}
    - \frac{1}{4}F_{2\mu\nu}F_2^{\mu\nu}
    - \frac{\delta}{2}F_{1\mu\nu}F_2^{\mu\nu},\nonumber\\
   {\cal{L}}_{\rm{St-mass}}& =& -\frac{1}{2} M_1^2  A_{1\mu}A_1^{\mu}
-\frac{1}{2} M_2^2 A_{2\mu}A_2^{\mu} - M_1M_2 A_{1\mu}A_2^{\mu},\nonumber\\
    \mathcal{L}_1 &=&
     J'_{\mu}A_1^{\mu}
    +J_{\mu}A_2^{\mu}.
       \label{stkt}
\eeqn
where $\mathcal{L}_{\rm kin}$ is the kinetic term,  $\mathcal{L}_{\rm St-mass}$ is the Stueckelberg mass mixing,
and  $\mathcal{L}_1$ is the interaction of the gauge field with the sources. Thus in
 Eq. (\ref{stkt})  $J_\mu$ is the current in the visible sector and $J_\mu'$ is the current in the hidden sector.
Let us assume that $M_1>> M_2$ and define the parameter $\epsilon= M_2/M_1$. 
In this case in the basis where both the kinetic energy and the mass terms are diagonalized, the 
interaction Lagrangian is given by 
\begin{eqnarray}
{\cal{L}}_{1} &= &\frac{1}{\sqrt{1-2\delta\epsilon+\epsilon^2}} \left( \frac{\epsilon-\delta}{\sqrt{1-\delta^2}} J_{\mu} +
       \frac{1-\delta\epsilon}{\sqrt{1-\delta^2}} J_{\mu}' \right) A_M^{\mu}
            \nonumber\\ & + &   \frac{1}{\sqrt{1-2\delta\epsilon+\epsilon^2}}
            \left(J_{\mu}- \epsilon J_{\mu}' \right)  A^{\mu}_{\gamma}.
            \label{kst}
\end{eqnarray}
Here $\delta$ parametrizes the kinetic mixing and $\epsilon$  parametrizes the Stueckelberg mass mixing.
In the above we identify $A^\mu_{\gamma}$ as the photon field, and $A^\mu_M$ as  the
massive  vector boson. Note that the photon field and the massive vector boson field
couple to both the visible sector current and 
the hidden sector current. We note that it is the parameter $\epsilon$ which is responsible for the generation of 
milli-charges for the fields in the hidden sector due to the coupling of the photon
to the hidden sector  which is proportional to $\epsilon$. For 
the case $\epsilon=0$, the photon does not couple to the hidden sector. However, the massive vector
boson couples to both the hidden sector and the visible sector through the kinetic mixing parameter $\delta$.

\subsection{Power spectrum of primordial perturbations and spectral indices \label{sec9.2}}
For slow roll inflation the power spectrum of curvature and tensor perturbations 
  ${\cal P_R}$ and ${\cal P}_t$ are given by    
   \begin{align}
   {\cal P_R}&= \frac{1}{12\pi^2}  \left(\frac{V^3}{M_{Pl}^6 V^{'2}}\right)_{k=RH}\,,\non   
   {\cal P}_t&= \frac{2}{3\pi^2} (\frac{V}{M_{Pl}^4})_{k=RH} \,. 
   \label{inflation-c.18q}
   \end{align} 
   where $V$ is the slow roll potential, $R$ is the cosmological scale factor and $H$ is the Hubble parameter 
   $H=\dot R/R$.   
One of the quantities of interest is the ratio $r$ defined by 
\begin{align}
r=  \frac{ {\cal P}_t(k_0) }{ {\cal P_R}(k_0) }\,,
\label{ratio-r}
\end{align}
where $k_0$ is a pivot scale. Also of interest are the spectral indices.  The curvature power spectrum for 
mode $k$ can be written in terms of the power spectrum for mode $k_0$ so that 
\begin{align}
{\cal P_R}(k) & = {\cal P_R}(k_0)  (\frac{k}{k_0})^{n_s(k) -1 }\,,
\label{ns}
\end{align}
where $n_s$ is the scalar spectral index.  For the tensor perturbations the power spectrum $P_t(k)$ can be expanded so that 
\begin{align}
{\cal P}_t(k) & = {\cal P}_t(k_0)  (\frac{k}{k_0})^{n_t(k)}\,,
\label{nt}
\end{align}
where $n_t$ is the tensor spectral index. 
   Slow roll inflation is often  described by  the parameters $\epsilon$ and $\eta$ which are  defined by   
\begin{align}
\epsilon
&= \frac{1}{2 } \left(\frac{M_{Pl} V'}{V}\right)^2, ~~\eta = \left|\frac{M^2_{Pl}V''(\phi)}{V(\phi)}\right|\,.
   \label{slow-indices}
\end{align}
The relation of these to spectral indices is given by Eq. (\ref{ns-nt-r}).

\subsection{Non-Gaussianity \label{sec9.3}} 
Non-Gaussianities are defined by three-point correlation functions of  perturbations involving 
three scalars, two scalars and a graviton, two gravitons and a scalar and three gravitons ~\cite{non-gauss}.
The dominant non-Gaussianity 
arises from the correlation function of three scalar perturbations. Thus for scalar perturbation $\zeta(\vec k)$ non-Gaussianity 
is defined by 
\begin{align}
< \zeta(\vec k_1) \zeta(\vec k_2) \zeta(\vec k_3)> = (2\pi)^7 \delta^3(\vec k_1+ \vec k_2+ \vec k_3)  \frac{\sum_i k_i^3}{\prod_i k_i^3} 
\left[ -\frac{3}{10} f_{NL} (P_k^{\zeta})^2 \right]\,,
\end{align}
where $P_k^{\zeta}$ is the scalar power spectrum and $f_{NL}$ is a measure of non-Gaussianity.

\subsection{Inflation in an axion landscape \label{sec9.5}}

We give here a short discussion of inflation in an axion landscape.  Let us consider
 $m$ pairs of chiral fields which are oppositely charged under a global $U(1)$ symmetry. 
We will call magnitude  of chiral fields saxions and the normalized phases axions.
We assume that there is  only one  $U(1)$ global symmetry under which the chiral fields  are charged and thus 
there is only one pseudo - Nambu - Goldstone - Boson (pNBG) which will act as the  inflaton. 
The $U(1)$ symmetry is broken by instanton effects and the potential can be decomposed
into into a fast roll involving $2m-1$ axions, and a slow roll which involves only the pNGB. 
The effective low energy theory for the pNGB is now  very different from the old natural
inflation model and one can generate inflation with $f< M_{Pl}$. We note here that the  
pNGB is not the axion of QCD but rather one that comes from  string theory.

To make the discussion more concrete let us  suppose  we have a set of fields $\Phi_i$ ($i=1, \cdots, m$) where $\Phi_i$ carry the same charge under the shift symmetry and the fields $\bar \Phi_i$ ($i=1, \cdots, m$) carry the opposite charge. 
  We may parametrize $\phi_k$ and $\bar \phi_k$ so that 
\begin{align}
\phi_k = (f_k + \rho_k) e^{ia_k/f_k}, ~~~\bar\phi_k = (\bar f_k + \bar \rho_k) e^{i\bar a_k/\bar f_k}\,.\nonumber
\end{align}
This allows us to write a non-trivial superpotential which can stabilize the saxions, i.e., 
\begin{align}
W= W_s (\Phi_i, \bar \Phi_i) + W_{sb}(\Phi_i, \bar \Phi_i)\,, \non
W_{sb}=\sum_{k = 1}^{m} \sum_{l = 1}^{q} {A}_{k  l} {\Phi}_{k}^{l} + \sum_{k = 1}^{m} \sum_{l = 1}^{q} {\bar{A}}_{k  l} {\bar{\Phi}}_{k}^{l}\,,
\label{All}
\end{align}
where $W_s$ is the symmetry preserving and $W_{sb}$ is the symmetry breaking superpotential. 
The stability conditions are given by 
\begin{align}
 W_{,\phi}=0= W_{,\bar\phi}\,.
\end{align}
The set up above has  $2m$  axionic fields  $a_1, \cdots, a_m, \bar a_1, \cdots, \bar a_m$.
 We can decompose this set into two: i.e., a set  that is invariant under the shift symmetry which consists of $b_k, \bar b_k, k=1,2,\cdots, m-1$ and 
 $b_+$ and the set which includes just one field $b_-$ which is sensitive to the shift symmetry. 
Here we exhibit only the field $b_-$ which acts as the inflaton~\cite{Nath:2017ihp}
\begin{align}
{b}_{-} = \frac{1}{\sqrt{\sum_{k = 1}^{m} {f}_{k}^{2} + \sum_{k = 1}^{m} {\overline{f}}_{k}^{2}}} \left( \sum_{k = 1}^{m} {f}_{k} {a}_{k} - \sum_{k = 1}^{m} {\overline{f}}_{k} {\overline{a}}_{k} \right)\,.
\end{align}
We can now decompose the potential into a slow roll part which involves only the field $b_-$ and a fast roll part which involves the rest of the fields.
The slow roll potential with stabilized saxions then takes the form 
\begin{align}
V(b) = V_{\text{fast}}+ V_{\text{slow}}(b_-)\,.\nonumber
\end{align}
Only the slow roll part of the potential, i.e., $V_{\text{slow}}$,  enters inflation~\cite{Nath:2017ihp}
\begin{align}
{V}_{\text{slow}}  =& \sum_{r = 1}^{q}   C_r { \left( 1 - \text{cos} \left( \frac{r}{ f_e}
{b}_{-}\right) \right)}
+ \sum_{l=1}^q \sum_{r=l+1}^q C_{rl} { \left( 1 - \text{cos} \left( \frac{r - l}{f_e } {b}_{-}\right) \right)}\,,
\label{slow}
\end{align}
where $C_r$ and $C_{rs}$ are determined in terms of $A_{kl}$ and $\bar A_{kl}$  in Eq.(\ref{All}) and 
\begin{align}
f_e= \sqrt{\sum_{k = 1}^{m} {f}_{k}^{2} + \sum_{k = 1}^{m} {\overline{f}}_{k}^{2} }\,.
\label{feffec}
\end{align}
A remarkable aspect of  ${V}_{\text{slow}}$ is that it depend only on an effective decay constant $f_{e}$. 
For $N$ number of fields and  $f_k=f=\bar f_k$, $  f_e= \sqrt{N} f$~\cite{Nath:2017ihp} (see also \cite{Ernst:2018bib})\,.
 The analysis of Eqs. (\ref{slow}) and (\ref{feffec})  imply  that we can get a 
 sub-Planckian domain of the axion decay constant by simply  enlarging the number of fields in the landscape. 
The $\sqrt N$ factor is as in  N-flation~\cite{Dimopoulos:2005ac}
although the mechanism of generation here  is very different. 

\subsection{Supersymmetric Dirac-Born-Infeld Inflation \label{sec9.6}}
Here we discuss the  simplest version of the supersymmetric  DBI Lagrangian involving just one chiral field. It is given by \cite{susy-dbi} 
\begin{equation}\label{DisplayFormulaNumbered:eq.dbiSuperLagrangian} 
	\mathcal{L}_{DBI}=\int d^4\theta \left(\Phi \Phi^\dagger+\frac{1}{16}\left(D^\alpha \Phi D_\alpha \Phi \right)\left({\overline{D}}^{\dot{\alpha}}\Phi^\dagger {\overline{D}}_{\dot{\alpha}}\Phi^\dagger \right)G\left(\Phi \right)\right)
	 +\int d^2\theta W\left(\Phi\right)+\int d^2\bar\theta
    W^* \left(\Phi^\dagger\right),
	\end{equation}
where $\Phi$ and $\Phi^\dagger$ are the chiral and anti-chiral superfields, $D_\alpha$ and ${\overline{D}}_{\dot{\alpha}}$ are the supercovariant derivatives, 
$T$ is a scale factor of the dimension of (mass)$^4$ where 

\begin{equation}\label{DisplayFormulaNumbered:eq.dbiG} 
	G\left(\Phi \right)=\frac{1}{T}\frac{1}{1+A\left(\Phi \right)+\sqrt{{\left(1+A\left(\Phi \right)\right)}^2-B\left(\Phi \right)}},
\end{equation}
\begin{equation}\label{DisplayFormulaNumbered:eq.dbiA} 
	A\left(\Phi \right)=\frac{\partial_\mu \Phi \partial^\mu \Phi^\dagger}{T},
	~~~B\left(\Phi \right)=\frac{\partial_\mu \Phi \partial^\mu \Phi \partial_\nu \Phi^\dagger \partial^\nu \Phi^\dagger}{T^2}.
\end{equation}
Ignoring fermions, the superfields have the expansion:

\begin{equation}
\begin{split}
	& \Phi \left(x,\theta,\overline{\theta}\right)
	%=\Phi_L\left(x,\theta,\overline{\theta}\right) \\
	%& 
	=\varphi \left(x\right)+\theta^\alpha \theta_\alpha F\left(x\right)+i \theta^\alpha {\sigma^\mu}_{\alpha \dot{\alpha}}{\overline{\theta}}^{\dot{\alpha}}\partial_\mu \varphi \left(x\right)+\frac{1}{4}\theta^\alpha \theta_\alpha {\overline{\theta}}_{\dot{\alpha}}{\overline{\theta}}^{\dot{\alpha}}\partial^\mu \partial_\mu \varphi \left(x\right),
\end{split}
\end{equation}
where $F$ obeys the cubic equation 
\begin{equation}\label{DisplayF_kormulaNumbered:eq.dbiF_kpqEquation} 
	F^3+p F+q=0.
\end{equation}
Here  $p$ and $q$ are given by 
\begin{align}
	p={\left(\frac{\partial W}{\partial \varphi}\right)}^{-1}\frac{\partial \overline{W}}{\partial \overline{\varphi}}\frac{1-2G \partial_\mu \varphi \partial^\mu 
	\overline{\varphi}}{2G},
	~~~q=\frac{1}{2G}{\left(\frac{\partial W}{\partial \varphi}\right)}^{-1}{\left(\frac{\partial \overline{W}}{\partial \overline{\varphi}}\right)}^2.
\end{align}
Integration on $\theta, \bar \theta$ in 
Eq. (\ref{DisplayFormulaNumbered:eq.dbiSuperLagrangian}) gives Eq. (\ref{DisplayFormulaNumbered:eq.dbiComponentLagrangianNoW})
 ~\cite{susy-dbi}
\begin{equation}\label{DisplayFormulaNumbered:eq.dbiComponentLagrangianNoW} 
\begin{split}
	 \mathcal{L}_{DBI} 
	& =-T\sqrt{1+2T^{-1}\partial_\mu \varphi \partial^\mu \overline{\varphi}+{T^{-2}\left(\partial_\mu \varphi \partial^\mu \overline{\varphi}\right)}^2-T^{-2} \left(\partial_\mu \varphi \partial^\mu \varphi \right)\left(\partial_\nu \overline{\varphi}\partial^\nu \overline{\varphi}\right)} \\
	& \indent{}+T+F \overline{F}+G\left(\varphi \right)\left(-2F\overline{F}\partial^\mu \varphi \partial_\mu \overline{\varphi}+F^2{\overline{F}}^2\right).
\end{split}
\end{equation}
As discussed in section \ref{sec4} a more realistic DBI model of inflation is given in ~\cite{Nath:2018xxe}
 based on two chiral fields
charged under a $U(1)$ global symmetry which is broken by instanton effects and non-Gaussianity is discussed consistent 
with PLANCK experimental constraints.

\end{document}